\documentclass[12pt,aaspp4,tighten,flushrt]{article}
\usepackage{euscript}
\def\ni{\noindent}
\def\.{\mathaccent 95}
\def\beq{\begin{equation}}
\def\ee{\end{equation}}

\def\be{\beta}

\def\ep{\epsilon}

\def\la{\lambda}

\def\frac#1#2{{\textstyle{{#1}\over {#2}}}}
\def\ni{\noindent}
\def\lsim{\mathrel{\rlap{\lower4pt\hbox{\hskip1pt$\sim$}}
    \raise1pt\hbox{$<$}}}
\def\gsim{\mathrel{\rlap{\lower4pt\hbox{\hskip1pt$\sim$}}
    \raise1pt\hbox{$>$}}}
\def\sqr#1#2{{\vcenter{\vbox{\hrule height.#2pt
         \hbox{\vrule width.#2pt height#1pt \kern#1pt
         \vrule width.#2pt}
         \hrule height.#2pt}}}}

\newbox\grsign \setbox\grsign=\hbox{$>$} \newdimen\grdimen \grdimen=\ht\grsign
\newbox\simlessbox \newbox\simgreatbox
\setbox\simgreatbox=\hbox{\raise.5ex\hbox{$>$}\llap
     {\lower.5ex\hbox{$\sim$}}}\ht1=\grdimen\dp1=0pt
\setbox\simlessbox=\hbox{\raise.5ex\hbox{$<$}\llap
     {\lower.5ex\hbox{$\sim$}}}\ht2=\grdimen\dp2=0pt

\def\x      {{\hbox{X-ray}}}

\def\etal   {{\it et~al.}}

\def\doublespace {\smallskipamount=6pt plus2pt minus2pt
                  \medskipamount=12pt plus4pt minus4pt
                  \bigskipamount=24pt plus8pt minus8pt
                  \normalbaselineskip=24pt plus0pt minus0pt
                  \normallineskip=2pt
                  \normallineskiplimit=0pt
                  \jot=6pt
                  {\def\smallskip {\vskip\smallskipamount}}
                  {\def\medskip   {\vskip\medskipamount}}
                  {\def\bigskip   {\vskip\bigskipamount}}
                  {\setbox\strutbox=\hbox{\vrule 
                    height17.0pt depth7.0pt width 0pt}}
                  \parskip 12.0pt
                  \normalbaselines}

\font\gkvec=cmmib10                         %for boldface lowercase
\def\bomega{\hbox{{\gkvec\char33}}}                  %bold face omega

\def\lb{\langle}
\def\rb{\rangle}

\def\bfJ{\bf J}
\def\bfA{\bf A}
\def\bbE{\overline {\bf E}}
\def\bbP{\overline {\bf P}}
\def\bw{\overline {\omega}}
\def\bv{\overline V}

\def\ts{\times}
\def\lb{\langle}
\def\rb{\rangle}
\def\curl{\nabla \ts}

\def\bbV{\overline {\bf V}}
\def\bfv{{\bf v}}

\def\bfV{{\bf V}}
\def\bfj{{\bf j}}

\def\bfe{{\bf e}}
\def\bfE{{\bf E}}
\def\bfw{{\bomega}}

\def\bfb{{\bf b}}

\def\bfB{{\bf B}}

\def\bbB{\overline{\bf B}}
\def\bbJ{\overline{\bf J}} 
\def\bbA{\overline{\bf A}} 
\def\bbE{\overline{\bf E}} 
\def\nb{\nabla}
\def\curl{\nb\ts}

\def\b0{b^{(0)}}
\def\v0{v^{(0)}}
\def\w0{\omega^{(0)}}
\def\bb0{\bfb^{(0)}}
\def\bv0{\bfv^{(0)}}
\def\bw0{\bfw^{(0)}}
\def\bj0{\bfj^{(0)}}

\def\ni{\noindent}
\begin{document}

\def\be{\begin{equation}}
\def\ee{\end{equation}}
\def\lab#1{\label{#1}}
\def\lrp#1{\left(#1\right)}
\def\BV{{\bf V}}
\def\etal{{\it et al.}\ }
\def\OV{\overline{\bf V}}
\def\E{{\bf E}}
\def\x{{\bf x}}

\def\pref#1{(\ref{#1})}
\def\beq{\begin{eqnarray}}
\def\eeq{\end{eqnarray}}
\def\nn{\nonumber}
\def\nt{\nabla\times}
\def\OE{\overline{\bf E}}
\def\A{{\bf A}}
\def\lra#1{\left\langle #1\right\rangle}
\def\bv{\bf v}
\def\OB{\overline{\bf B}}
\def\ove{\overline{E}}
\def\cnt{\cdot\nabla\times}
\def\b{{\bf b}}
\def\ob{\overline{B}}
\def\ao{\alpha\hbox{-}\Omega}
\def\B{{\bf B}}

\centerline {\bf Coronal activity as a consequence of 
dynamos in astrophysical rotators}
\medskip
\centerline{ Eric G. Blackman, Theoretical Astrophysics, Caltech 130-33 
Pasadena CA, 91125;} 
\centerline{Department of Physics \& Astronomy and Laboratory for
Laser Energetics}
\centerline {University of Rochester, Rochester, NY 14627}
\centerline {and}
\centerline {George B. Field, Harvard-Smithsonian 
Center for Astrophysics (CFA)}
\centerline {60 Garden St., Cambridge MA, 02139, USA} 

\medskip
\centerline{(submitted to MNRAS)}
\bigskip 
%\doublespace
\medskip
\centerline{\bf Abstract}
\medbreak

We show that the presence of a steady $\ao$ dynamo in astrophysical rotators
likely leads to an outflow of relative magnetic helicity 
and thus magnetic energy available for particle acceleration in a corona.
The connection between energy and magnetic helicity arises because
the generation of large scale field is linked to 
a generation of large scale magnetic helicity.
In a steady state, the relative magnetic 
helicity associated with the large scale field can
escape, accompanied by an equal and opposite small scale contribution
from the field, 
since the total magnetic helicity in large magnetic Reynolds number flows is 
conserved. From the helicity flow, a lower limit on the magnetic energy
deposited in  the corona can be estimated.  
Steady coronal activity and the dissipation of magnetic energy 
is therefore a signature of an internal dynamo. 
Our theoretical estimate of the power delivered by a mean field dynamo 
is consistent with that inferred from observations to be delivered to 
the solar corona, the Galactic corona, and Seyfert I AGN coronae.

%These result because the working dynamo necessarily requires
%finite but opposite flows of large and small scale magnetic helicity through 
%the host's surface boundary, and thus a flow of magnetic energy.  
%Specifically, we find that the nonthermal power $P$ of the object must
%satisfy the inequality,
%\be
%P >{1\over 3} \left|\lra{\alpha\ob^2}\right|R^2
%\ee
%where $\alpha$ is the turbulent dynamo coefficient, $\ob$ is the mean
%magnetic field, and $R$ is the radius of the object.  The Sun satisfies it by
%a factor of 30.  All astronomical objects whose large scale magnetic fields
%are produced by an $\ao$ dynamo, including stars, accretion disks, and
%galaxies must be shedding small scale magnetic helicity, and with it,
%nonthermal energy at at least the above rate.

\vfill
\eject

{\bf 1. Introduction}
\medskip

Understanding both the origin and destruction of magnetic fields 
is of fundamental importance to astrophysics. Not only are magnetic fields 
astrophysical entities in and of themselves, but they 
play an important intermediary role between gravitational
energy and radiation in rotating systems such as the Sun, galaxies, and
accretion discs. In this paper we will explore a link
between the origin of large scale magnetic fields from 
dynamo action in rotating systems and the export of magnetic energy 
into a corona which can dissipate and accelerate particles.

Large scale magnetic fields are observed in the Sun and
in spiral Galaxies.  The solar field changes on time scales much shorter 
than would be allowed  if the time scale of dissipation were
governed by resistivity alone.  The presence of a turbulent
solar convection zone leads naturally to the conclusion that 
an effective turbulent diffusivity must be at work. However,
for the field to maintain its strength in the presence
of turbulent diffusion, exponential amplification of the large
scale field must occur.
In the Galaxy the argument is similar--if the turbulent
ISM effectively diffuses magnetic field, then the large scale micro-gauss
fields must be somehow sustained.  
As the role of magnetic turbulence in accretion discs is thought to be fundamental for angular momentum transport, the rotating turbulent media
of accretion discs are also plausible sites for a similar mechanism. 

The leading, but controversial, candidate to explain
the origin of mean magnetic flux growth in stars and galaxies 
is mean-field turbulent magnetic dynamo theory 
(Moffatt 1978; Parker
1979; Krause \& R\"{a}dler 1980; Zeldovich et al.\ 1983, Ruzmaikin et al.\
1988, Beck et al.\ 1996). The theory appeals to a combination of helical
turbulence (leading to the $\alpha$ effect), differential rotation (the $\Omega$
effect),  and turbulent diffusion to exponentiate an initial seed mean
magnetic field. 
Steenbeck, Krause, and R\"adler (1966) developed a
formalism for describing Parker's (1955) concept that helical turbulence  
can twist toroidal fields  into the poloidal direction, where they can be
acted upon by differential rotation to regenerate a powerful large scale
magnetic field.  Their formalism involved breaking the total
magnetic field into 
%$\B$ into 
a large scale
component $\OB$ and a small scale component $\b$, and similarly for the
velocity field $\BV$.  They showed that $\OB$ satisfies the induction
equation
\be
{\partial\OB\over \partial t} = -c\curl\bbE,
\lab{2.4a} 
\ee
where
\be
\bbE=-\lrp{\OV\times \OB}/c  - \lb\bfv\ts\bfb\rb/c+
\lambda\nt\OB,
\lab{2.4aa} 
\ee
where the first term describes the effect of differential rotation
(``$\Omega$-effect"),
\be
\lb\bfv\ts\bfb\rb = \alpha\OB-\beta\nt \OB \lab{2.5a}
\ee
is the ``turbulent emf," and $\lambda=\eta c^2/4\pi$ is the magnetic
diffusivity.  Here $\alpha$ represents Parker's twisting (``$\alpha$ effect")
and $\beta(\gg \lambda)$ is the turbulent diffusivity. Steenbeck \etal
calculated $\OE$ to first order in $\ob$ and hence the dynamo
coefficients $\alpha$ and 
$\beta$ to zero order in $\ob$ from the statistics of the turbulence. 
 Recently Field, Blackman \& Chou (1999) have calculated $\alpha$ to all
orders in $\ob$.  When (\ref{2.4aa}) is substituted into (\ref{2.4a}), 
we have the mean-field dynamo equation:
\be
{\partial\OB\over \partial t} = \nt \lrp{\OV\times \OB} 
+\curl(\alpha \bbB)-\curl\curl(\beta +\la)\OB. \lab{6}
\ee
In the approximation that $\alpha$ and $\beta$ are independent of
$\ob$, (\ref{6}) 
is a linear equation for $\OB$ which can be solved as an eigenvalue
problem for the growing modes in the Sun and other bodies.
Boundary conditions play an important role in allowing net flux growth.
Also, a rapid growth of the fluctuating field necessarily accompanies the 
mean-field dynamo.  Its impact upon the growth of the mean
field, and the impact of the mean field itself on its own growth are
controversial.  

The controversy results because  
Lorentz forces from the growing magnetic field react back on and
complicate the turbulent  motions driving the field growth  (e.g. Cowling
1959, Piddington 1981, Kulsrud \& Anderson 1992; 
Kitchatinov et al 1994; Cattaneo \& Hughes 1996;
Vainshtein 1998; Field et al. 1999). It is tricky to
disentangle the back reaction of the mean field from that of the 
fluctuating field. Analytic studies and numerical
simulations seem to disagree as to the extent to which the dynamo
coefficients are suppressed by the back reaction of the mean field.

%{\bf 2. Dynamo Theory}

%The link is through   magnetic helicity, defined (Els\"asser
%1956) by
%\be
%H^M = \int_U \A\cdot \B\, d{\bf x} \lab{1}\ee
%where $\A$ is the vector potential of $\B=\nt \A$.  Field (1986) explains
%that $H^M$ is important in MHD  because under ideal conditions (vanishing
%resistivity) $H^M$ is conserved.  This has the consequence that in the $\ao$
%dynamo, in which kinetic helicity of the turbulence $\lb{\bv\cnt\bv}\rb$
%creates a large scale field carrying magnetic helicity, there has to be a
%compensating creation of small-scale field carrying 
%helicity of the opposite sign.

Pouquet \etal (1976) showed, from the numerical
solution of approximate equations describing the spectra of energy and
helicity in MHD turbulence, that the $\alpha$ effect conserves magnetic
helicity $(=\int({\bf A}\cdot{\bf B})d^3x)$, 
by pumping a positive (negative) amount to scales $>L$ (the outer
scale of the turbulence) while pumping a negative (positive) amount to
scales $\ll L$, where it is subject to Ohmic dissipation.  
They identified magnetic 
energy at the large scale with the $\OB$ of Steenbeck \etal (1966).  Thus,
dynamo action leading to an ever larger $\ob$, hence the creation of ever
more large scale helicity, can proceed as long as
small scale helicity of opposite sign can be dissipated by Ohmic diffusion.

The fate of small scale helicity is debated.
According to the nonlinear solutions of Pouquet \etal (1976), it cascades
to  large wave numbers where it is destroyed by Ohmic
dissipation.  According to several authors (Cattaneo \& Hughes 1996,
Gruzinov \& Diamond 1994, Seehafer 1994) the necessity for this process
limits the buildup of large scale helicity, and hence, large scale magnetic
fields. This would effectively eliminate the
$\ao$ dynamo as a practical process for creating large scale magnetic
fields in systems having a large magnetic Reynolds number $R_M$.

As part of an effort to investigate aspects of the backreaction problem
and the apparent differences between different simulation results,
Blackman \& Field (1999) have shown that when the scale of the averaging
is the scale of the simulation box, 
the coefficient $\alpha$ attains 
substantial values only 
if field gradients and non-periodic boundary conditions are present.  
This was shown to be related to
a flow of magnetic helicity through the boundary of the
system.  In the steady state, an equal and opposite flow of large
and small scale magnetic helicity should escape through the boundary
since total magnetic helicity is conserved in ideal MHD.
%Even if the required dissipation of small scale helicity
%were not taking place inside the object, the flow of small scale
%helicity through the boundary can drain this small scale helicity.
  
Differences in some apparently conflicting
simulations (Blackman \& Field 1999) may be interpreted to result from  
whether or not the boundary conditions are periodic (Cattaneo \& Hughes 1996)
or diffusive (Brandenburg \& Donner 1996).
Periodic boundary conditions can be used if the averaging scale
is significantly smaller than the size of the box (e.g. Pouquet et al. 1976;
Meneguzzi et al 1981; Balsara \& Pouquet 1999), which is not the case
in Cattaneo \& Hughes (1996).

The importance of boundary conditions makes it
natural to estimate the magnitude of quantities deposited 
through the boundary in a working dynamo.  In this regard, note that 
the Sun, the Galaxy, and accretion discs in AGN seem to harbor
steady active corona, 
requiring an energy source for heating or particle acceleration. 
Here we suggest that steady active coronae, in which
magnetic helicity and energy are deposited, 
naturally result when the underlying system harbors a mean field dynamo.  

In this paper we will estimate the helicity and thus magnetic energy flow into
a corona which accompanies a mean field magnetic dynamo 
The energy deposition rate is bounded below from the helicity
deposition rate (Field \& Blackman 1999).  The resulting estimates are
roughly consistent with that required of coronae in the Sun,
the Galaxy, and AGN accretion discs, all systems in which the operation
of a mean field dynamo is natural.  
The existence of and properties of a steadily active corona therefore 
provide some self-consistency checks and signatures of a working dynamo.

\medskip
\ni {\bf 2. Relative magnetic helicity flow and associated energy flux} 
\medskip

To explore the role of mean field gradients and boundary conditions
in determining the value of the $\alpha$ dynamo parameter, 
Blackman \& Field (1999) took Ohm's law
\be
\bfE= {-c^{-1}\bfV \times \bfB +\eta {\bf J}} 
%= -c^{-1}\lb{\bfv\times \bfb}\rb-{\bbV}\ts\bbB+ \eta \bbJ, 
\label{ohm}
\ee
and averaged the dot product with $\bfB$ to find
\be
\lb{\bfE \cdot \bfB}\rb 
%&-{1 \over 2}(\partial_\mu {\tilde h}^\mu +\partial_\mu {\overline h}^\mu)  
=\bbE\cdot \bbB+\lb{\bfe\cdot \bfb}\rb 
= -c^{-1}\lb{\bfv\times \bfb}\rb\cdot \bbB+\eta \bbJ\cdot \bbB+
\lb{\bfe\cdot \bfb}\rb 
%\frac12
%\part_\mu { {\overline h}^\mu},
\lab{mon21}
\ee
%where 
%\be
%\bbE= \lb{-c^{-1}\bfV \times \bfB +\eta {\bf J}}\rb 
%= -c^{-1}\lb{\bfv\times \bfb}\rb-{\bbV}\ts\bbB+ \eta \bbJ \lab{mon6}
%\ee
%for the case $\bbV=0$, 
where ${{\bf J}}$ is the current
density and $\eta$ is the resistivity.
%where we have used (\ref{mon6}).
%\beq
%\bbE \cdot \bbB= -c^{-1}\lra{\bfv\times \bfb}\cdot \bbB+\eta \bbJ \cdot
%\bbB\;, \lab{mon7}
%\ee
%which, from (\ref{2EMF}), shows that $\bbE\cdot \bbB$ is related to
%$\alpha$.  

A second expression for $\lb{\bf E}\cdot {\bf B}\rb$ also follows from 
Ohm's law without first splitting into mean and fluctuating components,
that is 
\be
\lb{\bfE\cdot \bfB}\rb 
%= \lb{-c^{-1} (\bfV\times \bfB )\cdot \bfB + \eta
=\eta \lra{\bfJ\cdot \bfB} = \eta \bbJ\cdot \bbB+\eta
\lb{\bfj \cdot \bfb}\rb =\eta\bbJ\cdot \bbB 
+c^{-1} \lambda \lra{\bfb\cdot\nabla\times \bfb}.
\lab{mon22}
\ee
Using  
(\ref{mon22}) and  (\ref{mon21}), Blackman \& Field (1999) obtain 
\be
\bbE\cdot\bbB= -c^{-1} \lb{\bfv\times \bfb}\rb \cdot \bbB 
 = c^{-1} \la \lra{\bfb\cdot
\nb\times \bfb } -\lb{\bf e}\cdot {\bf b}\rb
%\frac12 \part_\mu {{\overline h}^\mu}\; 
, \lab{mon23}
\ee
which they used to constrain  $\lb\bfv\ts\bfb\rb$.

Now consider ${\bf E}$ in terms of the vector 
and scalar potentials ${\bf A}$ and $\Phi$:
\begin{equation}
\bfE=-\nabla \Phi -(1/c)\partial_t {\bf A}.
\label{1}
\end{equation}
Dotting with $\bfB=\curl{\bf A}$ we have 
\begin{equation}
\bfE\cdot \bfB=- \nabla \Phi \cdot \bfB 
-(1/c){\bf B}\cdot \partial_t{\bf A}.
\label{02}
\ee
After straightforward algebraic manipulation, application of Reynolds rules 
(Ra\"dler 1980) and $\nabla\cdot \bfB=0$, this equation implies
\beq
\bfE\cdot \bfB
=- (1/2)\nabla\cdot   \Phi\bfB 
+(1/2)\nabla \cdot({\bf A}\ts {\bf E}) \nn\\
-(1/2c)\partial_t( 
{\bf A}\cdot \bfB) 
%=-(1/2c)\partial_t \int_U
%{\bf A}\cdot \bfB\ d^3x 
%=-\int_U \partial_t {H}^{0}d^3x 
%+\int_U \partial_i {H}^{i}
%=\int_U\partial_{0}
%{H}^{0}d^3x
=(-1/2c)\partial_\mu{H}^{\mu}\simeq 0,
\label{03a1},
\eeq
where $H^{\mu}=(H_{0},H_i)
=[{\bf A}\cdot \bfB, c\Phi\bfB 
-c{\bf A}\ts {\bf E}]$ 
is the magnetic helicity density 4-vector (Field 1986),
and the last similarity follows for nearly ideal MHD according to
(\ref{ohm}).

Taking the average of (\ref{03a1}) gives
\be
\partial_\mu{\overline H}^{\mu}=-2c\lb\bfE\cdot \bfB\rb=
-2c\bbE\cdot \bbB-2c\lb\bfe\cdot \bfb\rb\simeq 0
\label{03a2}.
\ee
Integrating (\ref{03a1}) over all of space, $U$, gives 
%
%\begin{equation}
%\int_U \bfE\cdot \bfB\ d^3x=-\int_U \nabla \Phi \cdot \bfB\ d^3x 
%-(1/c)\int_U{\bf B}\cdot \partial_t{\bf A}\ d^3x.
%\label{2}
%\end{equation}
%After straightforward algebraic manipulation, application of Reynolds rules 
%(R\"adler ) and $\nabla\cdot \bfB=0$, this equation implies
\beq
\int_U \bfE\cdot \bfB\ d^3x=- (1/2)\int_U\nabla\cdot   \Phi\bfB\ d^3x 
+(1/2)\int_U\nabla\cdot( {\bf A}\ts {\bf E})\ d^3x \nn\\
-(1/2c)\partial_t\int_U{\bf A}\cdot \bfB\ d^3x 
%=-(1/2c)\partial_t \int_U
%{\bf A}\cdot \bfB\ d^3x 
%=-\int_U \partial_t {H}^{0}d^3x 
%+\int_U \partial_i {H}^{i}
%=\int_U\partial_{0}
%{H}^{0}d^3x
=-(1/2c)\partial_t{\EuScript H}(\bfB)\simeq 0,
\label{3a1}
\eeq
where the divergence integrals vanish when converted to surface integrals.
The $\simeq$ follows for large $R_M$, and
we have defined the global magnetic helicity  
%4-vector (Field 1986) 
\begin{equation} 
{\EuScript H}(\bfB)
\equiv\int_U{\bf A}\cdot \bfB\ d^3x,
%[H_{0},H_i]=[(1/2c){\bf A}\cdot \bfB, (1/2)\nabla\cdot  ( \Phi\bfB) 
%-(1/2)\nabla \cdot ({\bf A}\ts {\bf E})].  
\label{3aa}
\end{equation}
where  $U$ allows for scales much larger than the mean field scales.
It is straightforward to show that a parallel argument 
%using (\ref{1}), (\ref{02}), and (\ref{3a1}) 
for the mean and fluctuating fields respectively leads to 
\be
\partial_t{\EuScript H}(\bbB)=
\partial_t\int_U\bbA\cdot\bbB\ d^3x
=-2c\int_U\bbE\cdot\bbB\ d^3x
\label{3aaa}
\ee
and
\be
\partial_t{\overline{\EuScript H}}(\bfb)=
\partial_t\int_U
\lb{\bf a}\cdot{\bf b}\rb\ d^3x
=-2c\int_U\lb\bfe\cdot\bfb\rb\ d^3x=-2c\int_U\bfe\cdot\bfb\ d^3x
=\partial_t{{\EuScript H}}(\bfb),
\label{3aab}
\ee
where the last two equalities in (\ref{3aab}) follow from 
the redundancy of averages; the 
volume integral amounts to averaging over a larger scale than
the inside brackets.

%If we take the average of (\ref{3a1})
%we then have 
%\be
%\partial_t{\overline{\EuScript H}}({\bf B})=
%\partial_t{\EuScript H}(\bbB)+\partial_t{{\bar {\EuScript H}}}
%(\bfb)=-2c
%\int_U \bbE\cdot \bbB\ d^3x-2c\int_U \bfe\cdot \bfb\ d^3x\simeq 0.
%\label{3ab}
%\ee
%Eqns. (\ref{3aaa}), (\ref{3aab}), and (\ref{3ab}) 
%show that a finite value of (\ref{mon23}) implies finite equal and opposite
%rates of change of global magnetic helicity from
%the small and large scale fields.
%Note that the time rate of change of the global helicities 
%are gauge invariant quantities.

%The exact same derivation carries through for both the 
%flucuating 
%[h_{0},h_i]=[(1/2c){\bf  a}\cdot \bfb, (1/2)\nabla\cdot  ( \phi\bfb) 
%-(1/2)\nabla \cdot ({\bf a}\ts {\bf e})].  
%\be
%\lb{\bf E}\cdot {\bf B}\rb = 
%{\overline{\bf E}}\cdot \bbB +\lb\bfe\cdot\bfb\rb
%= {1\over 2}
%\partial_\mu{H}^{\mu}={1\over 2}
%\partial_\mu{\widetilde h}^{\mu}+{1\over 2}\partial_\mu{\overline h}^{\mu}\simeq 0,
%\label{ref1}
%\ee

To estimate the energy flow implied by the  deposition of magnetic
helicity, we split  (\ref{3aaa}) and (\ref{3aab})  
into contributions from inside and outside the rotator.
One must exercise caution in doing so because ${\EuScript H}$
is gauge invariant, and hence physically meaningful, only 
if the volume $V$ over which ${\EuScript H}$ is integrated is bounded by 
a magnetic surface (i.e. normal component of $\bfB$ vanishes at the surface)
, whereas the surface separating the 
outside from the inside of the rotator is not magnetic in general.

Berger \& Field (1984) (see also Finn \& Antonsen 1985) 
showed how to construct a revised quantity they 
called relative helicity ${\EuScript H}_R$ which is gauge
invariant even if the boundary is not a magnetic surface.
They also showed that the total global helicity in a magnetically
bounded volume, divided into the sum of internal and
external regions, $U=U_{i}+U_{e}$ satisfies   
\be
{\EuScript H}_R(\bfB)={\EuScript H}_{R,o}(\bfB)+{\EuScript H}_{R,i}(\bfB).
\label{r4aaa}
\ee
Similar equations apply for $\bbB$ and $\bfb$, so (\ref{3aaa})
and (\ref{3aab}) can be written
\be
\partial_t{{\EuScript H}}_R(\bbB)
= \partial_t{{\EuScript H}}_{R,o}(\bbB) +\partial_t{{\EuScript H}}_{R,i}(\bbB),
\label{r4aab1}
\ee
and 
\be
\partial_t{{\EuScript H}}_{R}(\bfb)
=\partial_t{{\EuScript H}}_{R,o}(\bfb)
 +\partial_t{{\EuScript H}}_{R,i}(\bfb)
\label{r4aab2}
\ee
respectively. According to equation (62) of Berger \& Field (1984), 
\be
\partial_t{{\EuScript H}}_{R,i}(\bfB)=
-2c\int_{U_{i}}{\bfE\cdot\bfB} d^3x+2c\int_{D U_i}({\bfA}_p\ts {\bfE})\cdot d{\bf S},
\label{r4aac}
\ee
where $\bfA_p$ is the vector potential corresponding to a potential field ${\bf P}$ in $U_e$, and $DU_i$ indicates integration on the boundary surface of
the rotator.
Similarly, we have 
\be
\partial_t {{\EuScript H}}_{R,i}(\bbB)=
-2c\int_{U_{i}}{\bbE\cdot\bbB} d^3x
+2c\int_{D U_i}({\bbA}_p\ts 
{\bbE})\cdot d{\bf S}
\label{r4aad}
\ee
and
\be
\partial_t{{\EuScript H}}_{R,i}(\bfb)=
-2c\int_{U_{i}}{\lb\bfe\cdot\bfb\rb} d^3x+2c\int_{D U_i}\lb{\bf a}_p\ts 
{\bfe}\rb\cdot d{\bf S}.
\label{r4aae}
\ee
If we assume that the rotator is in a steady state
over the time scale of interest, then the left sides of 
(\ref{r4aad}) and (\ref{r4aae}) vanish.
The helicity supply rate, represented by the volume integrals
(second terms of (\ref{r4aad}) and (\ref{r4aae})),
are then equal to the integrated flux of relative magnetic 
helicity through the surface of the rotator.  Moreover, from 
(\ref{03a2}), we see that the integrated flux of the large scale 
scale relative helicity, $\equiv {\EuScript F}_{R,i}(\bbB)$,
and the integrated flux of small scale relative helicity,
$\equiv {\EuScript F}_{R,i}(\bfb)$,
are equal and opposite.  We thus have
\be
{\EuScript F}_{R,i}(\bbB)=
-2c\int_{U_{i}}{\bbE\cdot\bbB} d^3x
=2c\int_{U_{i}}{\bfe\cdot\bfb} d^3x=
-{\EuScript F}_{R,i}(\bfb).
\label{r4aaf}
\ee
To evaluate this,  we use  (\ref{2.4a}) and (\ref{2.4aa}) to
find
\be
\bbE=-c^{-1}(\alpha\bbB-\beta\curl\bbB),
\label{r4aaf}
\ee
throughout $U_i$.  
Thus  
\be
{\EuScript F}_{R,i}(\bbB)= -{\EuScript F}_{R,i}(\bfb)=
2 \int_{U_i}(\alpha\bbB^2-\beta\bbB\cdot\curl\bbB)d^3{\bf x}.
\label{r4aag}
\ee
This shows the relation between the 
equal and opposite large and small scale relative helicity
deposition rates and the dynamo coefficients.

Now according to Frisch \etal (1975), realizability of a helical magnetic
field requires its turbulent energy spectrum, $E^M_k$, to satisfy
\be
E^M_k(\bfb)\ge {1\over 8\pi} k|{{\EuScript H}}_k(\bfb)|\; , \lab{10}
\ee
where the equality applies to force-free fields with $\nt \B=\pm k\bf B$. 
%Equation (\ref{10}) is really just 
%the recognition that the helicity is a signed 
%quantity whereas energy is unsigned.
The same argument also applies to the mean field energy spectrum, so that 
\be
E^M_k(\bbB)\ge {1\over 8\pi} k|{{\EuScript H}}_k(\bbB)|. 
\lab{10a}
\ee
If we assume that the time and spatial dependences are separable
in both $E^M$ and ${{\EuScript H}}$, then 
a minimum power delivered to the corona can be derived.
For  the contribution from the small scale field, 
we have 
%is associated with
%${\dot {\EuScript H}}(\bfb)$ 
\beq
\dot E^M(\bfb) = 
\int {\dot E}^M_k(\bfb) dk \ge {1\over 8\pi}\int k|
{\EuScript F}_{k,R,i}(\bfb)| \, dk
\ge  {k_{\rm min}\over 8\pi} \int |{ {{\EuScript F}}_{k,R,i}(\bfb)| \, dk \ge 
{k_{\rm min}\over 8\pi} |{\EuScript F}_{R,i}({\bfb})}|\nn \\
={k_{\rm min}\over 8\pi} | {\EuScript F}_{R,i}(\bbB)|,
\lab{11}
\eeq  
where the last equality follows from the first equation in 
(\ref{r4aag}).
The last quantity is exactly the lower limit on 
$\dot E^M(\bbB)$.  Thus the sum of the lower limits on the total power 
delivered from large and small scales is 
${k_{\rm min}\over 8\pi} | {\EuScript F}_{R,i}(\bbB)|
+{k_{\rm min}\over 8\pi} | {\EuScript F}_{R,i}(\bfb)|
=2{k_{\rm min}\over 8\pi} | {{\EuScript F}}(\bbB)|$.  
Now for a mode to fit in the rotator, $k>k_{\rm min}=2\pi/H$, 
where $H$ is a characteristic scale height of the turbulent layer.  
Using (\ref{r4aag}), the total estimated energy delivered to the corona
(=the sum of the equal small and large scale contributions) is then
\beq
\dot E^M  \ge 2{k_{\rm min}\over 8\pi} |{\EuScript F}_{R,i}(\bfb)|
=2{k_{\rm min}\over 8\pi} |{\EuScript F}_{R,i}(\bbB)|
={{\rm V} \over H} \left| {\alpha\ob^2-\beta\bbB\cdot\curl\bbB}\right|,
\label{result}
\eeq
where $\rm V$ is the volume of the turbulent rotator. We will 
address implications of (\ref{result}) in the next sections.

\medskip
\ni {\bf 3. Relation to the Poynting Flux and the Force Free Case}
\medskip 

Blackman \& Field (1999) suggest that the combination of 
periodic boundary conditions and absence of mean field gradients does not 
allow a significant $\alpha$ when the
averaging scale is of order the scale of the periodic region.
Under these conditions, the last term on the right of 
(\ref{mon21}) was shown to vanish, and 
the main contribution to the turbulent EMF comes
from the second last term on the right.  This 
term is suppressed by the magnetic Reynolds number.
Thus, tests of $\alpha$ suppression when ignoring field gradients and
using periodic boundary conditions may be misleading as the apparent
suppression is not from the backreaction,  but is 
built in from the boundary conditions.
More generally, periodic boundary conditions are not appropriate for
characterizing mean magnetic flux growth unless there are many scale lengths
of the mean field within the simulation box of interest.

%To see that periodic boundary
%conditions are insufficient when the mean scale is of order the box
%size, note that 
%integrating the induction equation over a hemisphere
%of a rotator shows that to change the flux, there must be a finite
%$\alpha$ or $\beta$ contribution at the body's surfaces. 
%For quadrupole models, diffusion across the boundary is important 
%because there is no change in sign of the toroidal 
%mean field across the equatorial plane.  
%Thus mean poloidal loops which form from this field have the return field 
%opposite to that in the equator. 
%For dipole fields, the toroidal field vanishes at the equatorial
%plane, but not at the top surface. Periodic boundary conditions are
%again inappropriate for testing a fully working dynamo.

When field gradients are allowed (and when the
mean field scale is allowed to be smaller than the overall
scale of the system) the following conclusion
applies as a result of (\ref{result}) above:  
the flow of helicity through the boundary is required for 
substantial $\alpha^{(0)}$ unless the combination of
$\alpha^{(0)} \bbB^2-\beta^{(0)}\curl\bbB=0$.
Eqn. (\ref{result}) thus 
measures the extent to which a steady state field structure 
inside the object requires the deposition of magnetic helicity into the
corona, and gives a lower limit on the flow of magnetic energy 
to the exterior.  Since there is no physical
principle dictating that this quantity should be zero in general, 
we will later estimate the flow of helicity and 
magnetic energy through the boundary in the generic case for which 
the difference is represented by the order of magnitude of the first term.

%However, their calculations were based on a magnetic Reynolds number
%$R_M=30$, far smaller than typical of astronomical systems like the Sun. 
%We have shown that the above description is quantitatively
%accurate for low $R_M$, but if $R_M$ is large,
%dynamo action may be dramatically curtailed in a homogeneous system, as
%argued for example, by Seehafer (1994).  Instead, as shown by Blackman and
%Field (1999) dynamo action can occur in large $R_M$ systems only if the
%small scale helicity can escape to the surface of a body, something
%impossible in the homogeneous system considered by Seehafer.  

There is however, 
one exceptional case for which the difference in (\ref{result}) does vanish
exactly, and for which the energy deposition also vanishes exactly. This
is the case for which the field is force free. To see this, note that 
from Maxwell's equations we have
\beq
\partial_t{\bbB}=-c\curl {\bbE}.
\label{0.1}
\eeq
Dotting with $\bbB$ and using vector identities
gives Poynting's theorem (for $|B|>>|E|$)
\beq
(1/2)\partial_t{\bbB}^2=-c\nabla\cdot({\bbE}\ts{\bbB})
-c{\bbE}\cdot\curl{\bbB}.
\label{1}
\eeq
If we integrate over all of space, and assume a steady state
inside the object as we did for the helicity above,
we obtain
\beq
(1/2)\int_{U_e} \partial_t{\bbB}^2\ d^3x
=-c\int_{U_i}{\bbE}\cdot\curl{\bbB}\ d^3x-c\int_{U_e}{\bbE}\cdot\curl{\bbB}\ d^3x
,
\label{1.1}
\eeq
where the surface term vanished.
The internal contribution on the right side represents
a source term. Using (\ref{2.5a}) and the triple product rule,
we obtain for this term, represented as a energy depostion rate,  
\beq
{\dot E}^M|_{source}
%(1/2)\int_{U_e} \partial_t{\bbB}^2\ d^3x|_{\rm source} 
=\int_{U_i}\alpha \bbB\cdot\curl\bbB-\beta (\curl\bbB)^2
-c\int_{U_i}\bbV\cdot({\overline {\bf J}}\ts \bbB)\ d^3x.
\label{1.3}
\eeq
Now suppose we demand that the right side of (\ref{result}) vanishes.  
Then setting that right side equal to
zero gives $\alpha =\beta\bbB\cdot\curl\bbB/\bbB^2$.
Plugging this into back (\ref{1.3}) shows
that the right side of (\ref{1.3}) then vanishes completely
in the force (density) free case, ${\overline {\bf J}}\ts \bbB=0$.
Note that last term on the right of (\ref{1.3}) depends on the mean
velocity, which is not related to other terms on the right of
(\ref{1.3}) in any obvious way.  
It thus appears that the only natural case for which both helicity and 
and energy flux vanish exactly has to be force free.
But such a case is unphysical because there are no differentiable
field configurations for which ${\overline {\bf J}}\ts\bbB$=0
and $\bbB\ne 0$ when ${\overline {\bf J}}$ is  confined to a finite
volume (Moffatt 1978).  Because of this,  
even if ${\EuScript F_{U,i}}=0$, the energy deposition rate 
would not vanish in general, which is consistent with 
${\EuScript F_{U,i}}$, representing a lower limit as described
in the previous section.

The flow of relative magnetic helicity appears to be generically important for 
dynamo flux generation, but the exact value of the difference between 
terms in (\ref{result})
should depend on the solution for the dynamo equations for $\bbB$ and
the actual values of $\alpha$ and $\beta$, and the physics of the magnetic
diffusion and buoyancy.  The application of our result to
specific dynamo solutions, and a study of the boundary physics to 
see how buoyancy competes with turbulent diffusion in various environments
are both necessary components of future work.

In the next section, 
we discuss evidence that a significant
residual value of (\ref{result}) may be escaping into the
the coronae of the Sun, Galaxy and AGN accretion discs.

%{Escape of Small Scale Helicity from the Sun}
\medskip
\ni{\bf 4. Applications}
 
Keeping in mind the isssues addressed in the previous section, 
here we assume that the two terms on the right of (\ref{result}) 
do not cancel, and use the first term of (\ref{result}) as representative.

\medskip

\ni {\bf a. Solar Corona}

If we apply apply 
(\ref{result}) 
to each hemisphere of the Sun we have, using the first term as an
order of magntiude estimate
\beq
{\dot E}^M \gsim  \left({2\pi R_\odot^2 \over 3}\right)\alpha \bbB^2=
0.9 \ts 10^{28} \left({R \over 7\ts 10^{10}{\rm cm}}\right)^2 
\left({\alpha \over 40{\rm cm/s}}\right)
\left({{\overline B} \over  150 {\rm G}}\right)^2, 
\label{sun}
\eeq
where we have taken $\alpha\sim 40\,$cm~s$^{-1}$ from  Parker (1979),
and we have presumed a field in equipartition with kinetic motions
at of $150$G at a depth of $10$km beneath the solar surface in the convection
zone.

As this energy deposition rate is available for 
reconnection which can generate Alfv\'en waves and drive winds, and 
energize particles, we must compare this limit with the total of
downward heat conduction loss, radiative loss, and solar wind energy flux
in coronal holes, which cover $\sim 1/2$ the area of the Sun.  According to
Withbroe and Noyes (1977), this amounts to an approximately
steady activity of $2.5\times
10^{28}$erg~s$^{-1}$, about 3 times the predicted value of (\ref{sun}).

Other supporting evidence  
for deposition of magnetic energy and magnetic helicity in the sun 
includes: 
%as given at the July 1998, Chapman Conference entitled 
%``Magnetic Helicity in Space
%and Laboratory Plasmas" 
%(Pevtsov, Canfield \& Brown 1999)
%\begin{itemize}
(1)  The related pseudoscalar, current 
helicity $\lra{\B\cnt \B}$, has been 
been directly measured for flux tube filaments and their overlying
loop arcades penetrating the surface into the solar corona.  
The filaments, of order $10^3-10^4$km, have opposite sign of the
current helicity associated with the larger scale $10^5$km
overlying loops (Rust 1994; Rust and Kumar 1998; Martin 1998; Ruzmaikin 1999).
This small and large scale field having the opposite sign is
predicted above and is expected from Pouquet (1979).
%$R_\odot = 7\times 10^5$km.  
(2) The larger scale loops are associated with Coronal Mass Ejections (CMEs).  
%If an interplanetary spacecraft happens to pass through a
%CME, its magnetic field can be observed {\it in situ}.  
%Some CMEs carry current helicity and thus like  magnetic helicity.
Reverse ``S'' shaped CME loops dominate forward ``S'' shaped
CMEs with a 6 to 1 ratio in the northern hemisphere, and a similar
opposite ratio in the south (Rust \& Kumar 1998).
(3) There is a correlation between the sign of $\lra{\B\cnt\B}$ and
the sign of the observed twist of the field-aligned features in the
photo-sphere, implying that the parallel current responsible for the twist
originates below the photo-sphere and continues into the corona.
(4) Laboratory experiments indicate that twisted flux tubes are subject to kink
instability, leading to reconnection and magnetic energy release.
(5) The Yokoh satellite provides some of the 
most direct evidence for magnetic reconnection
in flares of various sizes (Masuda et al. 1994; Tsuenta 1996).
Shibata (1999) has a model in which
clouds of plasma (``plasmoids") are confined in twisted flux tubes which
reconnect with nearby flux, ejecting the plasmoid together with its
twisted flux, hence magnetic helicity.   Ejection of helicity is an important
part of the model.

In summary, currents along $\B$ twist emerging flux tubes, endowing them
with current helicity and therefore magnetic helicity as well.  Instability
leads to reconnection, allowing magnetic flux to escape in CMEs carrying
magnetic helicity.  The net result is that helical fields below the
photosphere escape the Sun, carrying magnetic helicity with them. 
Qualitatively, this is what is expected from Blackman \& Field (1999)
and shown explicitly in (\ref{result}).

{\bf b. Galactic Corona}

If we apply (\ref{result}) to the Galaxy, we have
for each hemisphere a lower limit on the luminosity delivered
to the corona, of 
\beq
{\dot E}^M \gsim \left({\pi R^2}\right)\alpha \ob^2=
\sim 10^{40} \left({R \over 12{\rm kpc}}\right)^2 \left({\alpha \over 10^5{\rm cm/s}}\right)
\left({\ob \over  5\ts 10^{-6}{\rm G}}\right)^2.
\label{gal}
\eeq
The value of $\alpha$ that we have scaled to 
is from Ruzmaikin et al. (1988).  Ferri\`ere (1993) 
suggests that $\alpha\sim 2 \ts 10^4$cm/sec at maximum,
which would lower the above estimate of the limit by a factor of 4. 
The study of Savage (1995)
suggests that the required steady energy 
input to the warm ionized medium in the
Galactic corona is $\sim 10^{41}$erg/sec, whereas that input into the highly
ionized coronal gas is $\sim 10^{40}$erg/sec. This compares
favorably with (\ref{gal}).
This is also implied by the study of Reynolds et al (1999), 
who argue that 
spatial variations of [S II]/H-Alpha and [N II]/H-Alpha line intensity 
ratios  in the halos of our and
other galaxies are inconsistent with pure photoionization models.
Instead, a  secondary heating mechanism is required that
increases the electron temperature at low densities $n_e$
with a dependence on $n_e$ to a lower power than the $n_e^2$ 
of photoionization.  Reynolds et al. (1999) estimate 
the required input heating rate of $4.1\ts 10^{40}$erg/s over a 12kpc radius. 
This again compares well with (\ref{gal}).

The observed tangled microGauss field in the halo also suggests
that magnetic reconnection and turbulent dissipation may be occurring.
(e.g. Beck et al. 1998)

{\bf c. Active Galactic Nuclei}

For a thin accretion disc in the turbulent viscosity framework,
we have roughly that the turbulent viscosity 
\beq
\beta\sim v_Tl \sim v_T^2/\Omega =\alpha_{ss}c_s H,
\label{ad0}
\eeq 
where $v_T$ is the turbulent velocity, $l$ is the typical
correlation scale, $\Omega$ is the rotational velocity,
$H$ is the height, $c_s$ is the sound speed, and $\alpha_{ss}$ is
the turbulent viscosity parameter (Shakura \& Sunyaev 1973).
If we assume that the angular momentum transport results from
MHD turbulence driven by a shearing instability (c.f. Balbus \& Hawley 1998),
then $v_T\simeq v_A$, 
in steady state, where $v_A$ is the Alv\'en speed. 
We thus have 
have $\alpha_{ss}\sim v_A^2/c_s^2$ where we have used the relation 
$\Omega H\sim c_s$, applicable to pressure supported discs.
Then using (\ref{ad0}), we also have 
\beq
\alpha_{ss}\simeq  (l/H)^{1/2}.
\label{ad01}
\eeq
Note next that the mass continuity equation for accretion discs
gives for the density 
\beq
\rho={{\dot M} \over 2\pi R H \alpha_{ss} v_r}
={L \over 2\pi 0.1 c^2  H^2  \alpha_{ss} c_s},
\label{ad01}
\eeq
where we have used $v_R=\alpha_{ss}Hc_s/R$ for the radial accretion speed, and
the accretion rate ${\dot M}=L/0.1 c^2$, where $L$ is the luminosity.
Using the above equations in (\ref{result}) we
have for each hemisphere
\beq
d_t E^M \gsim (\pi R^2)\alpha \bbB^2=
8\pi\alpha \alpha_{ss}c_s^2\pi R^2\rho
={2\pi\over 0.1 c^2}\alpha c_s L \left(R^2 \over H^2 \right).
\label{ad02}
\eeq
Now Brandenburg \& Donner (1996)  confirm 
the rough scaling of $\alpha \sim 2(l^2/H^2)\Omega H$.
Using this 
%and (\ref{ad01}) 
in (\ref{ad02}) gives
\beq
{\dot  E}^M \gsim  {8\pi \over 0.1}\alpha_{ss}{v_\phi^2 \over c^2}L
\sim 2 L \left({R_g \over R}\right) \left({\alpha_{ss} \over 0.01}\right),
\label{ad03}
\eeq
where $R_g$ is a gravitational radius.
This value compares well with the X-ray luminosities observed in 
Seyfert I galaxies (c.f. Mushotzsky et al. 1993), which can be of
order 30\% of the total luminosity, and variabilites indicate
emission from $R\sim 10R_g$.
The best working paradigm for the X-ray coronal luminosity
is the dissipation of magnetic energy in a non-thermal 
corona located above the disc 
(Galeev et al. 1979; Haardt \& Maraschi 1993; Field \& Rogers 1993)
which is consistent with the above result.

It is important to note that like the solar and Galactic coronae, 
the dissipation is required to be more or less steady.
Since dissipation is an 
exponentially decaying process, steady dissipation requires an exponential
feeding of this energy.  This is also consistent with the $\alpha-\Omega$ 
dynamo picture.

\medskip
\ni {\bf 5. Conclusions}
\medskip

The $\alpha-\Omega$ mean field dynamo in a turbulent rotator 
generically predicts some escape of relative large scale magnetic helicity 
and an equal and opposite escape of small scale relative magnetic 
helicity to the corona in the steady state. The helicity escape rate leads to 
a lower limit on the total magnetic energy deposition into the corona.
Exponential field growth from a dynamo would, in the steady state, lead
to a steady supply of magnetic energy into the corona which can feed a
steady non-thermal luminosity resulting from dissipation of magnetic fields. 
The estimated energy deposition rate 
agrees well with the nonthermal power from the Sun, Galaxy and Seyfert Is.
%As all of these sources are natural sites for $\alpha-\Omega$ type dynamos,
The steady flow of magnetic energy into  coronae thus provides an interesting
connection between $\alpha-\Omega$ dynamo and coronal dissipation paradigms
in a range of sources and provides a self-consistency check for dynamo 
operation. This deserves further scrutiny amidst the backreaction 
controversy debate.

Future work should consider specific dynamo solutions in different
settings to determine more precisely 
the predicted energy and helicity deposition rates 
from (\ref{result}) for a range of
dipole and quadrupole growth modes, and include a proper dynamical 
treatment of buoyancy. Also, the connection between 
observed oppositely signed large and small scale current helicities 
in e.g. the solar corona, needs to be correlated more precisely with
the predicted oppositely signed large and small scale magnetic helicities.

\pagebreak

\centerline{\bf References}

\def\item{\ni} 

\ni Balbus S.A \& Hawley J., Rev. Mod. Phys., 1998, 1.

\ni Blackman, E. G. \& Field, G. B. 1999, ApJ, submitted,
astro-ph/9903384.

\ni  Cattaneo, F., \& Hughes, D. W. 1996, Phys.\ Rev.\ E.\ {\bf 54}, 4532.

\ni Els\"asser, W. M. 1956, Rev.\ Mod.\ Phys.\ {\bf 23}, 135.

\ni Field, G. 1986, Magnetic Helicity in Astrophysics, in {\it
Magnetospheric Phenomena in Astrophysics}, R. Epstein \& W. Feldman, eds.\
AIP Conference Proceedings 144 (Los Alamos: Los Alamos Scientific
Laboratory 1986), 324.

\ni Balsara D. \& Pouquet A., Phys. of Plasmas, 1999, 6 89.

\ni Beck, R. et al., 1996,  Ann. 
Rev.  Astron. Astrophys., {\bf 34}, 155

\ni Berger M.A. \& Field G.B., 1984, JFM, 147 133.

\ni Ferri\`ere K. 1993, ApJ 404 162.

\ni  Field, G. B., Blackman, E. G., \& Chou, H. 1999, ApJ {\bf 513}, 638.

\ni Field G.B. \& Blackman E.G. 1999, in {\it Highly Energetic Physical Processes and Mechanism for Emisson from Astrophysical Plasmas}, IAU Symp.
195 eds. P.C.H. Martens and S. Tsuruta.

\ni Field G.B. \& Rogers. R.D., 1993, ApJ, 403, 94.

\ni Finn J., Antonsen T.M. 1985, Comm. Plasma Phys. and Contr. Fusion,9, 111.

\ni  Frisch, U., Pouquet, A., L\'eorat, J. \& Mazure, A. 1975, JFM {\bf
68}, 769.

\ni Galeev A.A.; Rosner R. Vaiana, G.S., 1979, ApJ., 229, 318. 

\ni Gruzinov, A. \& Diamond, P. H. 1994, PRL {\bf 72}, 1651.

\ni Haardt F. \& Maraschi, L., 1993, ApJ, 413, 507.

\ni Kitchatinov, L. L.,
Pipin V.V. and R{\"u}diger, G., \& Kuker, M. 1994, Astron. Nachr., 
{\bf 315}, 157

\ni Martin S., 1998, Solar Phys., 182, 107.
 
\ni Masuda S. et al., Nature, 1994, 371, 495.

\ni Meneguzzi, M. Frisch, U. Pouquet, A., PRL, 1981, 47, 1060.

\ni Mushotszky, R. Done, C., Ppunds K.A., 1993, ARAA, 31, 717. 

\ni  Parker, E. N., {\it Cosmical Magnetic Fields} (Oxford: Clarendon
Press).

\ni Parker, E. N. 1955, ApJ {\bf 122}, 293.

\ni Pevtsov, A., Canfield, R., \& Brown, X. eds.\ 1999, {\it Magnetic
Helicity in Space and Laboratory Plsamas}, (Amer.\ Geophys.\ Union).

\ni R{\"a}dler, K.-H. 1980, Astron. Nachr., {\bf 301}, 101

\ni Ruzmaikin, A., in 
{\it Magnetic Helicity in Space and Laboratory Plasmas}, 
Pevtsov, A., Canfield, R., \& Brown, X. eds. 1999, 
(Amer. Geophys. Union: Washington), p111.

\ni Ruzmaikin, A. A., 
Shukurov, A. M., and Sokoloff, D. D. 1988, {\sl Magnetic Fields of
Galaxies}, (Dordrecht: Kluver Press)

\ni Rust D.M., Geophys. Res. Lett., 1994, 21, 245.

\ni Rust D.M. \& Kumar A., 1994, ApJ, 464, L199.

\ni Pouquet, A., Frisch, U., \& Leorat, J. 1976, JFM {\bf 77}, 321.

\ni Reynolds R.J., Haffner L.M., Tufte S.L., 1999, ApJ, 525, L21.

\ni Seehafer, N. 1994, Europhys.\ Lett.\ {\bf 27}, 353.

\ni Shakura N.I. \& Sunyaev R.A., 1973, A\& A., 24, 337.

\ni  Shibata, K. 1999, in {\it Magnetic Helicity in Space and
Laboratory Plasmas}, A. Pevtsov, R. Canfield, and X. Brown eds.\ (American
Geophysical Union, 1999).

\ni Steenbeck, M., Krause, F., \& R\"adler, K. H. 1966, Z. Naturforsch.
{\bf 21a}, 369.

\ni Tsuneta S., 1996, ApJ 456, 840.

\ni Vainshtein S., 1998, PRL, 80, 4879.

\ni Withbroe, G. L. \& Noyes, R. W. 1977, Ann.\ Rev.\ Astron.\
Astrophys.\ {\bf 15}, 363.

\end{document}